\documentclass[12pt,preprint,graphicx]{aastex}

\slugcomment{Version: 2009 January 15}

\lefthead{Hachisuka et al.}
\righthead{Distance to outer Galaxy star}

\begin{document}

\title{The distance to a star forming region in the Outer arm of the Galaxy}
   \author{K. Hachisuka\altaffilmark{1,2},
           A. Brunthaler\altaffilmark{2},
           K. M. Menten\altaffilmark{2},
           M. J. Reid\altaffilmark{3}, \\
       Y. Hagiwara\altaffilmark{4}
           and N. Mochizuki\altaffilmark{5}
}

\altaffiltext{1}{Key Laboratory for Research in Galaxies and
Cosmology, Shanghai Astronomical Observatory, Chinese Academy of
                 Science, Shanghai, 200030, China}
\altaffiltext{2}{Max-Planck-Institut f\"ur Radioastronomie,
                  Auf dem H\"ugel 69, Bonn, 53121, Germany}
\altaffiltext{3}{Harvard-Smithsonian Center for Astrophysics,
              60 Garden Street, Cambridge, MA 02138, USA}
\altaffiltext{4}{National Astronomical Observatory, Mitaka, Tokyo 181-8588,
                 Japan}
\altaffiltext{5}{Institute of Space and Astronautical Science, Japan Aerospace
                 Exploration Agency, 3-1-1 Yoshinodai, Sagamihara, Kanagawa,
                 229-8510, Japan}

\begin{abstract}

We performed astrometric observations with the VLBA of WB89$-$437, an H$_2$O
maser source in the Outer spiral arm  of the Galaxy.  We measure an annual
parallax of 0.167$\pm$0.006 mas, corresponding to a heliocentric distance of
6.0$\pm$0.2 kpc or a Galactocentric distance of 13.4 $\pm$ 0.2 kpc.
This value for the heliocentric distance is considerably smaller than the
kinematic distance of 8.6 kpc.
This confirms the presence of a faint Outer arm toward $l=135^{\circ}$.
We also measured the full space motion of the object and find a large peculiar
motion of $\sim$20 km\,s$^{-1}$ toward the Galactic center. This peculiar
motion explains the large error in the kinematic distance estimate. We also
find that WB89$-$437 has the same rotation speed as the LSR, providing more
evidence for a flat rotation curve and thus the presence of dark matter in
the outer Galaxy.

\end{abstract}

\keywords{astrometry, masers, stars: distances,
          Galaxy: structure, kinematics and dynamics}

\section{Introduction}
\begin{figure}[ht]
\centering
\resizebox{\hsize}{!}{\includegraphics[angle=-90]{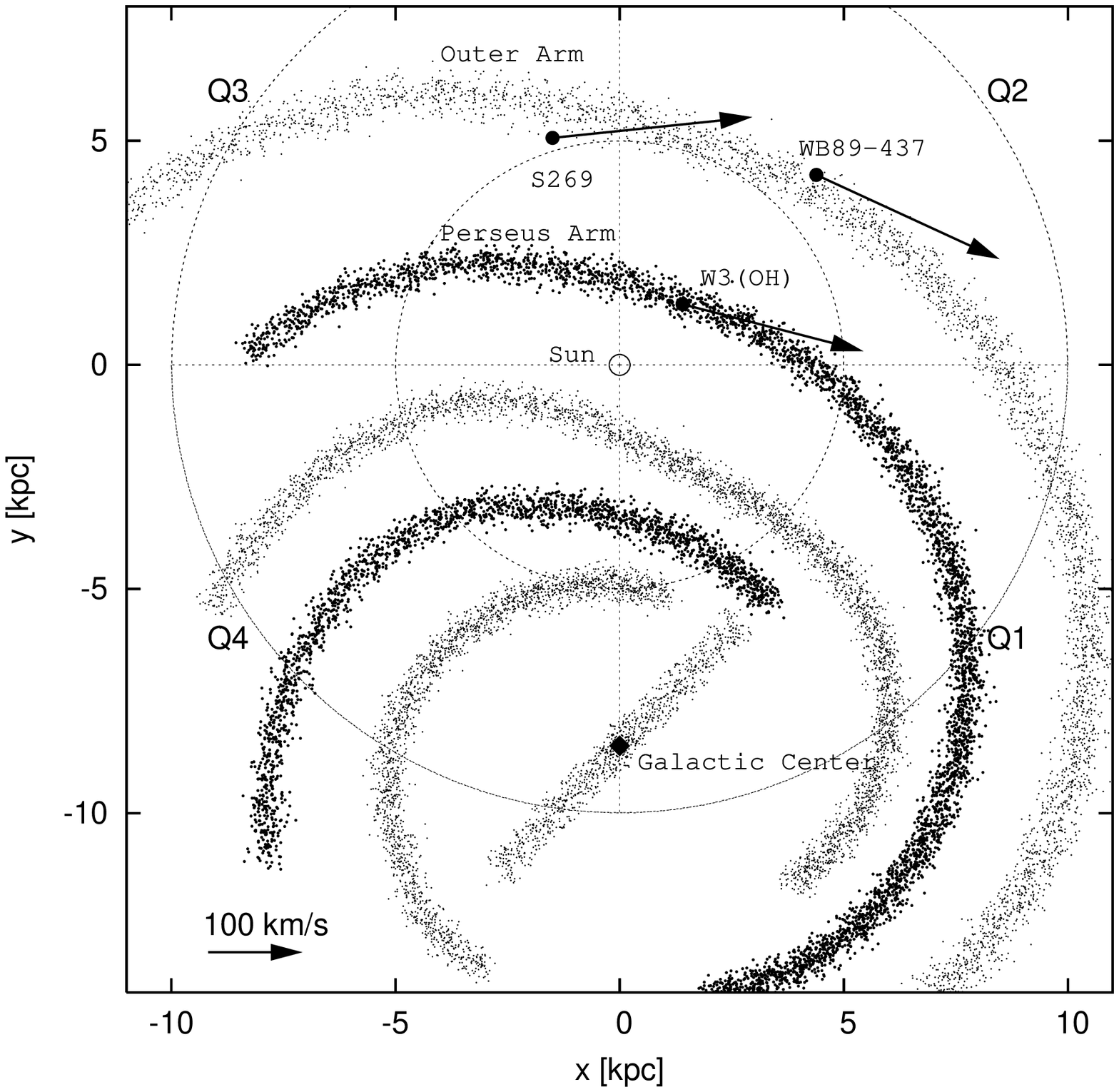}}
\caption{Schematic view of the spiral arms of the Milky Way
(Georgelin \& Georgelin 1976) with the positions of W3OH (Xu et
al. 2006; Hachisuka et al. 2006), S269 (Honma et al. 2007),
WB89-437 (this paper), and their motions relative to the Galactic
Center (arrows). The IAU value R$_0$=8.5 kpc was assumed. Two dark
arms represent principal arms, namely the Perseus and
Scutum-Centaurus arm (Benjamin 2008). Also shown is the location
of the central bar from Benjamin et al. (2005). The labels Q1-Q4
designate the standard four quadrants of the Galaxy.} \label{mw}
\end{figure}

The Outer spiral arm of the Milky Way (also called the Cygnus arm) is
located beyond the Perseus arm and may be the outermost
arm of the Galaxy's stellar disk (see Fig.~\ref{mw}).
It is traceable across portions of the fourth and first quadrants of
the Galaxy and it can be detected through the emission of HI
and molecular gas (Nakanishi \& Sofue 2003; Nakanishi \& Sofue 2006).
The arm may continue through the second and third quadrants, as shown
by spectroscopic and photometric observations of open clusters (Pandey et al.
2006) and HII regions (Russeil et al. 2007). Indeed, Honma et al. (2007) used
the VERA array and measured the annual parallax and proper motion of H$_2$O
masers in S\,269, a star forming region at a Galactic longitude of
196$^\circ$. They found a Galactocentric distance of 13.6 kpc
(for R$_0$=8.5\,kpc), consistent with it being in the Outer arm.

The size and structure of our Galaxy have been mainly determined by kinematic
distances, which involve comparing measured radial velocities with a kinematic
model (e.g., G\'omez 2006). However, kinematic distances depend
strongly on the Galactic rotation model chosen and peculiar motions render
them questionable (e.g., Xu et al. 2006). This is particularly true for the
outer Galaxy, since the Galactic rotation speed is quite uncertain
(e.g. Brand \& Blitz 1993).

Recently, distances of Galactic maser sources have been directly determined
by annual parallax measurements that use phase-referencing VLBI techniques
(e.g. Hachisuka et al. 2006; Xu et al. 2006; Honma et al. 2007).
Such parallax measurements can yield distances of many kiloparsecs with
errors less than ten percent, allowing \textit{direct} distance determinations
for objects located near the edge of the stellar disk of our Galaxy.

Wouterloot et al. (1993) searched for H$_2$O maser sources toward IRAS sources
with $\mbox{}^{12}$CO emission and found several for which their kinematic
distances indicated large distances from the Galactic center. Most of these
maser sources seem to be located in the Perseus arm, but several sources
have large negative LSR velocities and are probably in the Outer arm
(Wouterloot et al. 1995).

One of these sources, WB89$-$437 (IRAS 02395+6244), is located at
a Galactic longitude of $l=135^\circ$ and has a kinematic distance
of 8.6 kpc, which would place it beyond the Outer arm. It is one
of the most luminous FIR source with the strongest H$_2$O maser
emission in the outer part of the Galaxy. Since radio continuum
emission has not been detected, the spectral type of the exciting
star is thought to be B1 (Rudolph et al. 1996) or later (Brand et
al. 2001). The young stellar object WB89$-$437 itself is deeply
embedded and, testifying for its youth, drives an outflow traced
by high velocity CO and H$_{2}$O masers. The latter are found
offset by roughly 3 arcseconds from the position of the peak of CO
and CS emission.

Because of its strength and location in the outer part of the Galaxy,
we selected this maser as the target for an annual parallax measurement.
Here we report the results of these measurements, which yielded the distance
and proper motion of this star forming region in the outer part of the Galaxy.

\section{Observation and data reduction}
We used the NRAO\footnote{The National Radio Astronomy Observatory is a
facility of the National Science Foundation operated under cooperative
agreement by Associated Universities, Inc.} Very Long Baseline Array (VLBA)
to observe WB89$-$437 under program BH136.
The observations involved rapid switching between an extragalactic background
continuum radio source and the H$_2$O maser. The observations were performed at
5 epochs spread over about 10 months in 2006 (see Table~\ref{table:1} for
details). The separations between adjacent epochs were typically 3 months
 and the observations were performed close to the dates of the maximum and
minimum of the annual parallax signature in right ascension and declination.
We limited the observing time span to 10 months, instead of 1 year, to
minimize the effects of maser variability.

We observed two 16 MHz wide bands with one band centered on the maser's
centroid LSR velocity of $-70.0$ km~s$^{-1}$. The data were correlated with
1024 spectral channels in each band, resulting in a channel spacing of
0.21 km~s$^{-1}$. We used the ICRF extragalactic radio source J0244+6228 at
R.A.=02$^h$44$^m$57.696849$\pm$0.000168$^s$  and
Dec.=+62$^\circ$28$'$06.51470$\pm$0.00097$''$ (J2000) (Fey et al. 2004)
as the phase-reference source.  This source had already been used for the
parallax observation of the H$_2$O maser in W3OH (Hachisuka et al. 2006).
J0244+6228 had no detectable extended structure and its high flux density,
near 1~Jy at 22 GHz, makes it an excellent source for phase-referencing VLBI.
Moreover, its angular separation from WB89$-$437 is only 0$^{\circ}$.5,
promising high quality relative astrometry. We switched between maser and
reference source every 30 seconds, yielding $\approx$22 seconds on-source
time per scan. At each epoch, we obtained total on-source integration times
of $\approx50$ minutes for each source. The total observing time was spread
over 6 hours, including the calibrator observations, the slewing time of
the telescopes, and observations of a second maser which will be presented
elsewhere.

In order to correct for small zenith delay errors in the atmospheric model of
the VLBA correlator (see Reid \& Brunthaler 2004), we performed
{\it geodetic-like} observations at 22 GHz (Brunthaler et al. 2005) at the
beginning and end of each VLBA observation.  These observations involved 22
ICRF sources with positions accurate to better than 1 mas and sampled a wide
range of source elevations with a frequency setup involving eight 4\,MHz bands
at left circular polarization that spanned a frequency range of 450\,MHz. The
multi-band delays and rates from a fringe fit to these quasars were then
fit with a model that consisted of a zenith delay offset at all antennas
as well as a clock offset at all antennas except the reference antenna.

Most of the calibration and data reduction were carried out with standard
procedures for spectral-line observations using the NRAO's Astronomical Image
Processing System (AIPS). First, we applied the latest values of the Earth's
orientation parameters and corrected for effects of the changing
feed parallactic angles. Next, we removed atmospheric zenith delay errors,
corrected for voltage offset in the samplers, and applied antenna gains and
system temperatures for the amplitude calibration.
Next, we corrected for electronic delay and phase differences among the
IF bands using data from NRAO\,150. Many H$_2$O maser spots could be imaged
after calibration.
We determined their peak positions on images by fitting elliptical Gaussian
components and traced individual maser spots over the five epochs.

\section{Results}
\begin{figure}[ht]
\centering
\includegraphics[height=12cm]{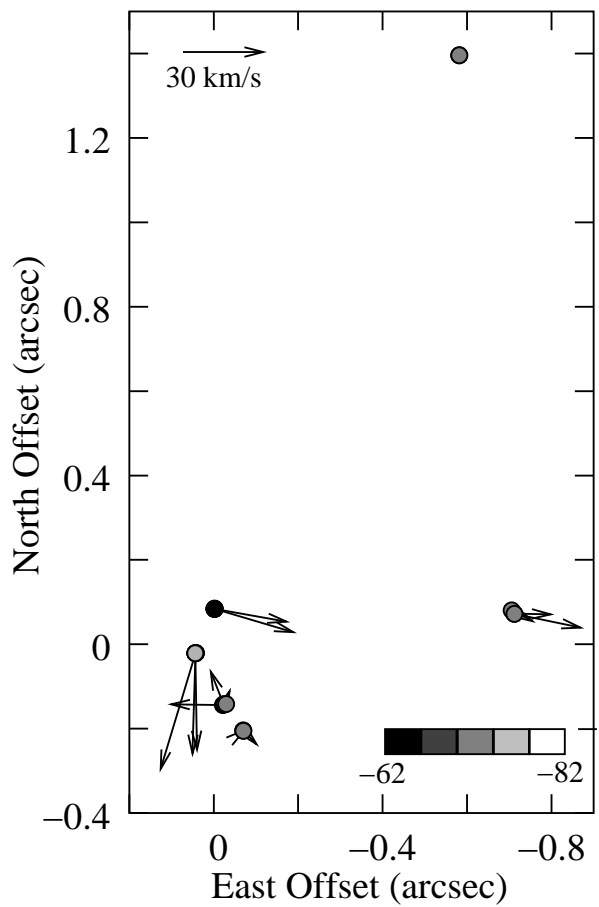}
\caption{Spot distribution (circles) and internal motions not
affected by fringe-rate smearing (arrows) relative to a positional
reference feature (maser spot 2 with a LSR velocity of -73.16
km~s$^{-1}$ in Table 2.) of maser components that were detected at
two or more epochs. The coordinate of the origin of the map is
02$^h$43$^m$28.5680$^s$, 62$^\circ$57$'$08.388$''$ (J2000). The
gray scale indicates radial velocities as coded in the panel at
the lower right. The radial velocity of thermally emitting
molecular lines is $-$72 km~s$^{-1}$. } \label{map}
\end{figure}

Twenty H$_2$O maser features were detected in the LSR velocity range $-62$ to
$-76$ km~s$^{-1}$, distributed over an area of 0$''$.8 $\times$ 1$''$.6
(Fig. ~\ref{map}). 16 of the 20 H$_2$O maser features were detected at 2 or
more epochs and 4 of them were detected at all 5 epochs. We used three of
the four H$_2$O maser features that were detected at
all five epochs for the parallax fitting. The last feature was located
at a distance of $\sim$1.4$''$ from the phase center, which is larger than
our nominal field of view ($\sim$0.7$''$). Hence, it was heavily affected
by fringe-rate smearing and was not used for the parallax estimate.

\subsection{Annual parallax and distance}

One maser feature was detected in thirty channels, and two features were
detected over seven frequency channels. First, we fitted the parallaxes and
proper motions to data from each spectral channel individually.  The residuals
to the fits showed reduced $\chi^2$ values that were relatively large
($\sim$ 2--8), since the formal position errors from the Gaussian fits to the
maps underestimate the true position uncertainty.
The true position error of a maser spot relative to a background source is
usually dominated by systematics, owing to uncalibrated residual atmospheric
delays and/or maser spot structural variations.
Hence, we added ``error floors'' in quadrature to the formal position
uncertainties until we achieved $\chi^2$ per degree of freedom values of near
unity for the data from each coordinate.  For this dataset we needed error
floors between 10 and 60 $\mu$as for different maser features.
The individual parallaxes and proper motions for each feature are shown in
Table~2 together with their formal errors from the fit. Also given
are the average parallaxes and proper motions for each feature together with
the standard error of the mean and the standard deviation (in parentheses).

\begin{figure}[ht]
\centering
\resizebox{0.7\hsize}{!}{\includegraphics[angle=0]{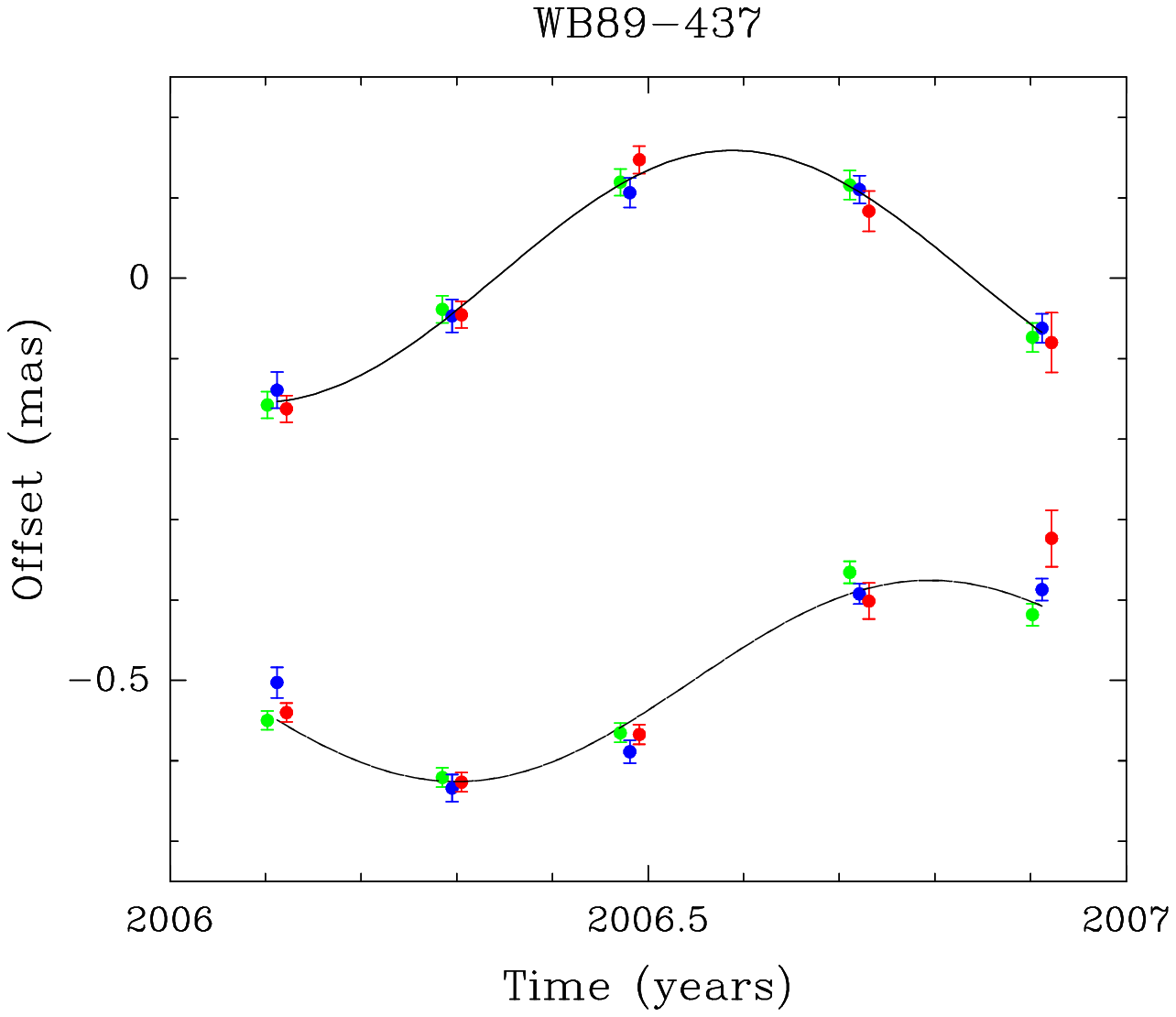}}
\resizebox{0.7\hsize}{!}{\includegraphics[angle=0]{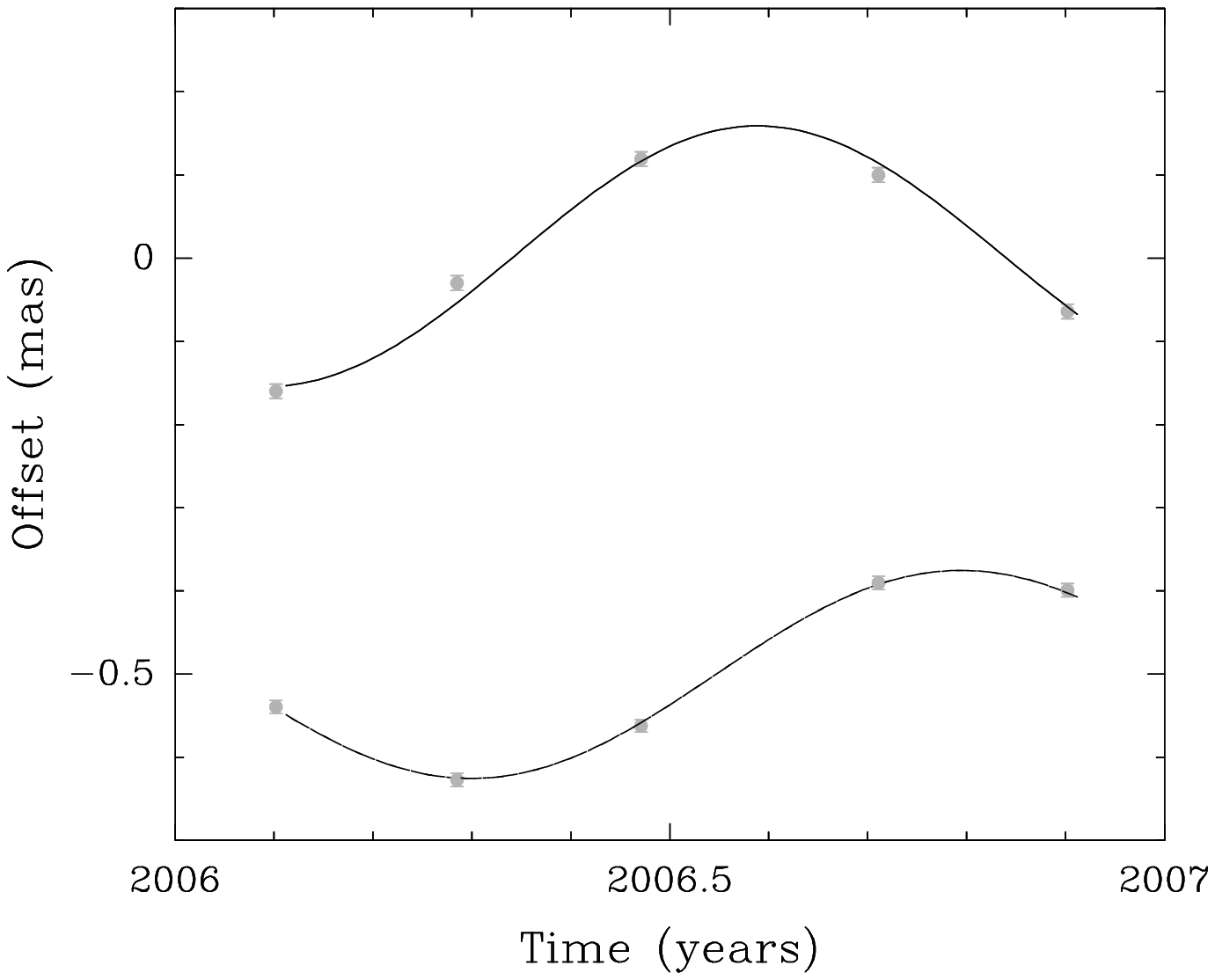}}
\caption{{\bf Top:} Annual parallax signal of one channel (i.e.
the first channel) from each maser feature in WB89$-$437. The
individual proper motions and position offsets are removed. The
data for the different maser features are slightly shifted in time
for clarity (green: maser feature 1, blue: feature 2, red: feature
3). The upper data shows the parallax in east-west direction while
the lower data shows the parallax in north-south direction. {\bf
Bottom:} Annual parallax signal for the average positions of the
data from all channels from the three maser spots after removing
the individual proper motions and position offsets. The upper data
shows the parallax in east-west direction while the lower data
shows the parallax in north-south direction. } \label{parallax}
\end{figure}

Next, we fitted all channels from any given maser feature simultaneously with
one parallax but different proper motions and position offsets for each
channel. The resulting parallaxes for each maser feature are also given in
Table~2.  All three features have consistent parallaxes of
0.164$\pm$0.0014, 0.163$\pm$0.004 mas and 0.164$\pm$0.007 mas.
We also performed a combined parallax fit with all channels from
the three features. The resulting parallax is 0.164$\pm$0.0016 mas. For this
fit we used error floors of 17 and 12 $\mu$as in right ascension and
declination, respectively. Fig.~\ref{parallax} (top)  shows the parallax
signal for one channel (i.e. the first channel) from each feature and
the combined fit to all channels. The individual proper motions and position
offsets were removed before plotting.

Combining the results of several maser spots can lead to underestimation
of the parallax uncertainty, since the measurements may not be statistically
independent. Random-like errors (e.g. from map noise and possibly maser spot
structure variations) will not be correlated among different maser spots.
However, systematic errors (e.g. caused by residual atmospheric delay errors)
will affect all maser spots in one epoch in a very similar way.  The most
conservative approach would be to assume 100\% correlation and multiply the
formal error by $\sqrt{N}$, where $N$ is the number of individual data sets
used in the fit. This would give an error of $\sqrt{44}\times0.0016$ mas
= 0.011 mas.

To estimate the effect of the systematic errors on our parallax
measurement, we examined position residuals after removing the
proper motions and position offsets (e.g. the data plotted in
Fig.~\ref{parallax}, top) and calculated the average maser
position of all channels from all three features in each epoch. The
averaging should reduce the random error, but leave the systematic
error unaffected. A parallax fit to the averaged data points
(Fig.~\ref{parallax}, bottom) yielded a value of
0.167$\pm$0.003\,mas, which is consistent with the value from the combined
fit. Here we used error floors of 12 and 5 $\mu$as in
right ascension and declination, respectively, to achieve $\chi^2$
values of $\sim$1. This suggests that the systematic errors are larger in
right ascension, contrary to our previous experience, where systematic errors
in declination dominate. However, because the source is at high declination,
systematic errors in right ascension may dominate in such a
case for some geometrical configurations (Pradel et al. 2006).
Furthermore, the 8.6 GHz image of the calibrator J0244+6228
in the VLBA calibrator tool shows a weak jet mainly directed toward the east.
Thus, small changes in this jet could introduce larger systematic errors
in right ascension than in declination. The different error floors are also
the reason why the parallax fit to the averaged data is slightly larger than
the parallax fits to the individual features.

In order to allow for errors induced by the VLBI structure of the
reference quasar, we multiply our final error by a factor of two
and obtain 0.167$\pm$0.006\,mas, which we adopt for the parallax
of WB89$-$437. This parallax corresponds to a distance of
$6.0\pm0.2$\,kpc from the Sun, or a Galactocentric distance of
$13.4\pm0.2$\,kpc (assuming R$_0$=8.5\,kpc).

\subsection{Proper motion and full space motion in the Galaxy}

Since we have a distance, proper motion and radial velocity, we
can determine the 3-dimensional motion of WB89$-$437 in the
Galaxy. The absolute proper motion of a Galactic object relative
to an extragalactic source depends not only on the annual parallax
but also on differential Galactic rotation, the Solar Motion and
the peculiar motion (relative to Galactic rotation) of the object.
In addition, H$_2$O masers in star forming regions usually
participate in outflows with typical velocities of a few tens of
km~s$^{-1}$.  Therefore, the absolute proper motions and radial
velocities of maser spots can contain a significant component from
internal motions.

The absolute proper motions of the three maser features range from $-$1.18 to
$-$1.35 mas~year$^{-1}$ in right ascension and from 0.59 to 1.05 mas~year$^{-1}$
in declination. The average motion of these
features is $-$1.27 $\pm$ 0.05 mas~year$^{-1}$ in right ascension and 0.82
$\pm$ 0.13 mas~year$^{-1}$ in declination. However, it is not clear that
the average proper motion of the three maser spots represent the true
motion of the whole object.

To evaluate the effect of relative internal motions, we calculated
the average motion of 11 maser features relative to a reference
feature (maser spot 2 with an LSR velocity of $-73.16$
km~s$^{-1}$). Here we used all maser spots which were detected in
at least two epochs and not affected by fringe-rate smearing. The
results are shown in Fig.~2 and Table~3. Also given is the average
proper motion together with the standard error of the mean and the
standard deviation (in parentheses). The average motion is -0.02
$\pm$ 0.15 mas~year$^{-1}$ in right ascension and -0.35 $\pm$ 0.18
mas~year$^{-1}$ in declination. The true proper motion of the
whole source can then be estimated by the sum of the average
motion and the absolute proper motion of the reference feature.
This gives a total motion of -1.22 $\pm$ 0.30 mas~year$^{-1}$ in
right ascension and 0.46 $\pm$ 0.36 mas~year$^{-1}$ in
declination. Here we also multiplied the final error by a factor
of two to allow for errors induced by the background quasar.

We assume IAU values for the distance of the Sun from the Galactic
center of R$_0$=8.5~kpc and the circular rotation speed of the
LSR of $\Theta_0$=220 km\,s$^{-1}$.  We also adopt the Solar Motion
with respect to the LSR in km\,s$^{-1}$ from HIPPARCOS data,
(U,V,W) = ($10.00\pm0.36, 5.25\pm0.62, 7.17\pm0.38$) (Dehnen \& Binney 1998).
Then our distance measurement of 6.0$\pm$0.2\,kpc, the proper motion of
$-$1.22$\pm$0.30 mas\,year$^{-1}$ in right ascension and
0.46$\pm$0.36 mas\,year$^{-1}$ in declination, and a radial velocity of
$-$72$\pm$2 km\,s$^{-1}$ from CO and CS measurements (Brand et al. 2001),
which should be close to the stellar velocity,
implies a peculiar motion relative to a circular Galactic rotation of:

\begin{eqnarray}
U'&=&23.1  \pm  4.1\,{\mathrm {km\,s}}^{-1},\nonumber\\
V'&=&-3.6 \pm  7.9\,{\mathrm {km\,s}}^{-1},\nonumber\\
W'&=&0.8  \pm 9.9\,{\mathrm {km\,s}}^{-1}.\nonumber
\end{eqnarray}

\noindent
Here, $U'$ denotes the velocity component toward the Galactic Center, $V'$ is
the component in direction of Galactic rotation, and $W'$ is the component
toward the North Galactic Pole.

If one uses a different rotation model of the Milky Way with
R$_0$=8.0\,kpc and $\Theta_0$=236\,km\,s$^{-1}$, which is consistent with
the measured proper motion of the Galactic center super-massive black hole,
Sgr A* (Reid \& Brunthaler 2004), we get

\begin{eqnarray}
U'&=&15.2  \pm  4.1\,{\mathrm {km\,s}}^{-1},\nonumber\\
V'&=&-4.2 \pm  8.0\,{\mathrm {km\,s}}^{-1},\nonumber\\
W'&=&0.8  \pm 9.9\,{\mathrm {km\,s}}^{-1}.\nonumber
\end{eqnarray}

Hence, WB89$-$437 rotates in both cases with the approximately same velocity
as the LSR but shows a large peculiar motion of $\approx20$ km~s$^{-1}$ toward
the Galactic center.

\section{Discussion}

\subsection{WB89$-$437 in the outer Galaxy}

The Galactocentric distance of WB89$-$437 is 13.4~kpc (for
R$_0$=8.5\,kpc), which places the source well outside the Perseus
spiral arm, which is at a Galactocentric distance of $\approx10.0$
kpc in this direction (Xu et al. 2006, Hachisuka et al. 2006).
While the distribution of molecular gas in the outer Galaxy shows
that the Outer arm becomes much weaker near $l=90^{\circ}$
(Nakanishi \& Sofue 2006), some CO clouds with high radial
velocities are distributed from $l=131^{\circ}$ to $137^{\circ}$
(Digel et al. 1996). This indicates that WB89$-$437 (at
$l=135^{\circ}$) is located in a weak extension of the Outer arm.
Potential arm objects at $l>137^{\circ}$ have been identified
(Russeil et al. 2007) which suggest that this faint arm continues
to the third Galactic quadrant. This view is supported by Honma et
al. (2007) who used the VERA array and measured the annual
parallax and proper motion of a H$_2$O maser at a Galactic
longitude of 196$^\circ$. They found a Galactocentric distance of
13.6 kpc (for R$_0$=8.5\,kpc), consistent with it being in the
Outer arm. There are many maser sources at the outer Galaxy whose
annual parallax have not been measured.  With direct measurements
of distances to these sources, we may understand the structure of
the Outer (and the Perseus) arm in more detail.

With a Galactic latitude of $b=2.8^\circ$ and a distance of 6 kpc, WB89-437
is located $\sim$300 pc above the Galactic plane. With our upper limit of 10
km\,s$^{-1}$ for the motion perpendicular to the plane, it would have needed
more than 30 million years to reach its current position.
This is much larger than the age of this object.
Thus, WB89-437 must have formed already far above the plane.

\subsection{Galactic dynamics in the Outer arm}

The annual parallax distance to WB89$-$437 (6.0$\pm$0.2 kpc) is significantly
smaller than its kinematic distance of 8.6$\pm$0.4 kpc, assuming recommended
IAU values R$_0$ and $\Theta_0$. However, we note that our measured distance
is in slightly better agreement with the kinematic distance of 7.3$\pm$0.3 kpc
that one obtains for R$_0$=8.0\,kpc and $\Theta_0$=236\,km\,s$^{-1}$
(Reid \& Brunthaler 2004), but there is still a significant discrepancy.
The large peculiar motion of WB89$-$437 of $\sim$20 km\,s$^{-1}$ toward
the Galactic center has a component of $\sim$10 km\,s$^{-1}$ toward the Sun and
is partially responsible for this discrepancy.

This is similar to the case of W3OH (Xu et al. 2006 ;  Hachisuka et al. 2006),
which is located at a similar Galactic longitude but is in the Perseus arm.
W3OH also shows a large peculiar motion and a true distance which is smaller
than the kinematic distance.
Anomalous motions in the Perseus arm are well known to exist (Humphreys 1978);
they could be caused by spiral density waves.
On the other hand, since the outer arm in the Galaxy is a faint arm, one
expects that the influence of a density wave on it is smaller than on an
inner arm. In fact, spiral structure of HI gas at large Galactic radii
($>$25 kpc) is not seen (Levine et al. 2006).

A flat rotation curve of the outer Galaxy implies the existence of dark
matter since the density of visible matter in the outer Galaxy is smaller
than in the inner Galaxy. The measured circular orbital speed of WB89$-$437 is
consistent with that of the LSR. This confirms the result of Honma et al.
(2007) that the rotation curve of the Milky Way is constant out to
$\sim$13.5 kpc and provides more solid evidence for the existence of dark
matter in the outer region of the Galaxy. However, since massive star forming
regions can have large peculiar motions, it may be possible, although unlikely,
 that the rotation  curve falls by $\sim20$ km\,s$^{-1}$ between the Sun and
the Outer arm, but the source has a compensating component in the direction
of Galactic rotation.

\section{Conclusion}
We performed astrometric VLBA observations toward the Galactic H$_2$O maser
source WB89$-$437 in the Outer arm of the Galaxy. The measured annual parallax
of 0.167$\pm$0.006 mas corresponds to a heliocentric distance of 6.0$\pm0.2$
kpc (a Galactocentric distance of 13.4 $\pm$ 0.2 kpc). This confirms the
presence of a faint Outer arm in the direction of $l=135^{\circ}$. Our
measured  distance is smaller than the kinematic distance of 8.6 kpc.
We also estimate the 3D motion of the object with respect to a Galactic
reference frame and find that the discrepancy between kinematic distance and
true distance is caused, in part, by a large  peculiar motion of
$\sim$20 km\,s$^{-1}$ toward the Galactic center. We also find that WB89-437
has the same rotation speed as the LSR, confirming a flat rotation curve in
the outer Galaxy.

\acknowledgments
This work was supported in part by the National
Natural Science Foundation of China (grants 10573029, 10625314,
10633010 and 10821302) and the Knowledge Innovation Program of the
Chinese Academy of Sciences (Grant No. KJCX2-YW-T03), and the
National Key Basic Research Development Program of China (No.
2007CB815405). KH acknowledges the support by China Postdoctoral
Science Foundation (grant 20070410745). Andreas Brunthaler was
supported by the DFG Priority Programme 1177.

\clearpage

\begin{deluxetable}{cccc}
\label{table:1}
\tablenum{1}
\tablewidth{0pt}
\tablecaption{Details of the VLBA observations \label{table:1}}
\tablehead{
\colhead{Date} & \colhead{J0244+6228} & \colhead{Beam} &
\colhead{Image rms} \nl
\colhead{} & \colhead{Flux density} & \colhead{HPBW; P.A} &
\colhead{per channel} \nl
\colhead{} & \colhead{(mJy)}& \colhead{(mas$^2$); ($^\circ$)}&
\colhead{(mJy\,bm$^{-1}$)}}
\startdata
 2006 Feb 09 &  443 &  0.79$\times$0.28; 11 & 12 \nl
 2006 Apr 17 &  758 &  0.80$\times$0.27; 12 & 12 \nl
 2006 Jun 24 &  852 &  0.80$\times$0.28; 11 & 16 \nl
 2006 Sep 20 & 1244 &  0.80$\times$0.28; 10 & 17 \nl
 2006 Nov 29 & 1116 &  0.81$\times$0.28; 12 & 13 \nl
\enddata
\end{deluxetable}

\begin{deluxetable}{llrrccc}
\tablenum{2}
\tablewidth{0pt}
\label{table:2}
\tabletypesize{\scriptsize} \tablecaption{Annual parallax and
absolute proper motion of the spots which appeared in all five
epochs. Coordinates are relative to the phase center
(02$^h$43$^m$28.5680$^s$, 62$^\circ$57$'$08.388$''$) at epoch
2006.5. Column five lists the parallax estimates. Columns six and
seven give the motion on the plane of the sky along the right
ascension and declination, respectively. $\mu_{\rm RA}$ is the
true RA motion multiplied by cos(Dec.).} \tablehead{
\colhead{Feature} & \colhead{V$_{\rm LSR}$} & \colhead{$\Delta$RA}
& \colhead{$\Delta$Dec} & \colhead{$\pi$} & \colhead{$\mu_{\rm
RA}$} & \colhead{$ \mu_{\rm Dec}$}  \nl \colhead{}  &
\colhead{(km\,s$^{-1}$)} & \colhead{(mas)} & \colhead{(mas)} &
\colhead{(mas)} & \colhead{(mas\,year$^{-1}$)} &
\colhead{(mas\,year$^{-1}$)}} \startdata 1 &  -66.42   &   -69.34
&   -203.72   & 0.168$\pm$0.006 & $-$1.36$\pm$0.02 & 0.58$\pm$0.03
\nl 1 &  -66.63   &   -69.34  &   -203.72   & 0.180$\pm$0.009 &
$-$1.37$\pm$0.02 &  0.60$\pm$0.02 \nl 1 & -66.84   &   -69.35  &
-203.72   & 0.156$\pm$0.008 & $-$1.34$\pm$0.02 &  0.63$\pm$0.03
\nl 1 &  -67.05   &   -69.36  & -203.73   & 0.160$\pm$0.006 &
$-$1.36$\pm$0.02 &  0.56$\pm$0.03 \nl 1 &  -67.26   &   -69.36  &
-203.73   & 0.176$\pm$0.003 & $-$1.37$\pm$0.03 &  0.57$\pm$0.01
\nl 1 &  -67.47   &   -69.37  & -203.72   & 0.162$\pm$0.009 &
$-$1.36$\pm$0.02 &  0.62$\pm$0.03 \nl 1 &  -67.68   &   -69.38  &
-203.72   & 0.170$\pm$0.006 & $-$1.37$\pm$0.02 &  0.61$\pm$0.02
\nl 1 &  -67.89   &   -69.39  & -203.73   & 0.166$\pm$0.007 &
$-$1.37$\pm$0.02 &  0.59$\pm$0.02 \nl 1 &  -68.10   &   -69.40  &
-203.72   & 0.161$\pm$0.009 & $-$1.36$\pm$0.02 &  0.61$\pm$0.03
\nl 1 &  -68.31   &   -69.40  & -203.73   & 0.165$\pm$0.007 &
$-$1.36$\pm$0.02 &  0.59$\pm$0.02 \nl 1 &  -68.53   &   -69.41  &
-203.73   & 0.170$\pm$0.007 & $-$1.37$\pm$0.02 &  0.58$\pm$0.02
\nl 1 &  -68.74   &   -69.41  & -203.74   & 0.166$\pm$0.006 &
$-$1.36$\pm$0.02 &  0.57$\pm$0.01 \nl 1 &  -68.95   &   -69.42  &
-203.74   & 0.168$\pm$0.006 & $-$1.37$\pm$0.02 &  0.55$\pm$0.01
\nl 1 &  -69.16   &   -69.43  & -203.74   & 0.164$\pm$0.009 &
$-$1.36$\pm$0.03 &  0.56$\pm$0.02 \nl 1 &  -69.37   &   -69.43  &
-203.74   & 0.164$\pm$0.009 & $-$1.35$\pm$0.03 &  0.57$\pm$0.02
\nl 1 &  -69.58   &   -69.43  & -203.74   & 0.176$\pm$0.004 &
$-$1.37$\pm$0.03 &  0.56$\pm$0.01 \nl 1 &  -69.79   &   -69.44  &
-203.74   & 0.170$\pm$0.006 & $-$1.36$\pm$0.02 &  0.57$\pm$0.01
\nl 1 &  -70.00   &   -69.44  & -203.74   & 0.170$\pm$0.006 &
$-$1.35$\pm$0.02 &  0.58$\pm$0.01 \nl 1 &  -70.21   &   -69.45  &
-203.74   & 0.173$\pm$0.004 & $-$1.35$\pm$0.03 &  0.59$\pm$0.01
\nl 1 &  -70.42   &   -69.45  & -203.74   & 0.163$\pm$0.006 &
$-$1.34$\pm$0.02 &  0.61$\pm$0.02 \nl 1 &  -70.63   &   -69.45  &
-203.74   & 0.167$\pm$0.004 & $-$1.35$\pm$0.02 &  0.60$\pm$0.01
\nl 1 &  -70.84   &   -69.45  & -203.74   & 0.166$\pm$0.006 &
$-$1.35$\pm$0.02 &  0.60$\pm$0.02 \nl 1 &  -71.05   &   -69.45  &
-203.75   & 0.166$\pm$0.002 & $-$1.34$\pm$0.02 &  0.61$\pm$0.01
\nl 1 &  -71.26   &   -69.46  & -203.75   & 0.160$\pm$0.007 &
$-$1.33$\pm$0.02 &  0.63$\pm$0.02 \nl 1 &  -71.47   &   -69.46  &
-203.75   & 0.156$\pm$0.009 & $-$1.33$\pm$0.02 &  0.63$\pm$0.03
\nl 1 &  -71.68   &   -69.47  & -203.75   & 0.156$\pm$0.008 &
$-$1.33$\pm$0.02 &  0.62$\pm$0.02 \nl 1 &  -71.90   &   -69.47  &
-203.76   & 0.153$\pm$0.010 & $-$1.32$\pm$0.02 &  0.63$\pm$0.03
\nl 1 &  -72.11   &   -69.48  & -203.77   & 0.148$\pm$0.011 &
$-$1.32$\pm$0.03 &  0.60$\pm$0.05 \nl 1 &  -72.32   &   -69.49  &
-203.76   & 0.176$\pm$0.002 & $-$1.36$\pm$0.03 &  0.60$\pm$0.01
\nl 1 &  -72.53   &   -69.49  & -203.77   & 0.168$\pm$0.007 &
$-$1.34$\pm$0.03 &  0.58$\pm$0.02 \nl \hline
&\multicolumn{2}{l}{Average}  & &  0.166$\pm$0.001 (0.007) &
$-$1.35$\pm$0.004 (0.02)& 0.59$\pm$0.004 (0.02)\nl
&\multicolumn{2}{l}{Combined fit}  & & 0.164$\pm$0.0014 \nl \hline
2 &  -72.53   &   -28.65    &  -141.63   & 0.155$\pm$0.007 &
$-$1.25$\pm$0.02 &  0.79$\pm$0.07 \nl 2 & -72.74 &   -28.67 &
-141.68   & 0.152$\pm$0.008 & $-$1.17$\pm$0.02 & 0.83$\pm$0.03 \nl
2 &  -72.95   &   -28.72 &  -141.71   & 0.176$\pm$0.006 &
$-$1.18$\pm$0.03 &  0.80$\pm$0.01 \nl 2 & -73.16 &   -28.77 &
-141.74   & 0.176$\pm$0.005 & $-$1.20$\pm$0.03 & 0.81$\pm$0.01 \nl
2 &  -73.37   &   -28.82 & -141.76   & 0.173$\pm$0.005 &
$-$1.21$\pm$0.02 &  0.79$\pm$0.01 \nl 2 & -73.58 &   -28.86 &
-141.77   & 0.162$\pm$0.002 & $-$1.17$\pm$0.03 & 0.82$\pm$0.01 \nl
2 &  -73.79   &   -28.88 & -141.78   & 0.151$\pm$0.010 &
$-$1.11$\pm$0.02 &  0.84$\pm$0.06 \nl \hline
&\multicolumn{2}{l}{Average}  & &      0.164$\pm$0.004 (0.011) &
$-$1.18$\pm$0.02 (0.04)& 0.81$\pm$0.01 (0.02)\nl
&\multicolumn{2}{l}{Combined fit}  & & 0.163$\pm$0.004\nl \hline 3
&  -74.63   &   -24.93    &  -141.84   & 0.191$\pm$0.012 &
$-$1.39$\pm$0.04 &  1.06$\pm$0.07 \nl 3 & -74.85 &   -24.88    &
-141.85   & 0.158$\pm$0.016 & $-$1.20$\pm$0.04 & 1.07$\pm$0.04 \nl
3 &  -75.06   &   -24.86 &  -141.87   & 0.150$\pm$0.014 &
$-$1.18$\pm$0.04 &  1.05$\pm$0.04 \nl 3 & -75.27 &   -24.86    &
-141.87   & 0.158$\pm$0.015 & $-$1.22$\pm$0.04 & 1.05$\pm$0.04 \nl
3 &  -75.48   &   -24.85 & -141.87   & 0.170$\pm$0.017 &
$-$1.29$\pm$0.05 &  1.03$\pm$0.04 \nl 3 & -75.69 &   -24.87    &
-141.87   & 0.174$\pm$0.022 & $-$1.36$\pm$0.07 & 1.02$\pm$0.05 \nl
3 &  -75.90   &   -24.89 & -141.87   & 0.140$\pm$0.014 &
$-$1.39$\pm$0.07 &  1.07$\pm$0.03 \nl \hline
&\multicolumn{2}{l}{Average}  & &      0.163$\pm$0.006 (0.016) &
$-$1.29$\pm$0.03 (0.08)& 1.05$\pm$0.01 (0.02)\nl
&\multicolumn{2}{l}{Combined fit}  & & 0.164$\pm$0.007\nl \hline
\multicolumn{2}{l}{Average} & & &      0.164$\pm$0.001 (0.0015)  &
$-$1.27$\pm$0.05 (0.09) & 0.82$\pm$0.13 (0.23)\nl
\multicolumn{2}{l}{Combined fit all} & & & 0.164$\pm$0.0016\nl
\hline
\enddata
\end{deluxetable}

\begin{deluxetable}{crrrrr}
\label{table:3}
\tablenum{3}
\tablewidth{0pt}
\tablecaption{Positions and internal proper motions of all detected maser
features relative to the reference feature 2.}
\tablehead{
\colhead{Feature} &\colhead{V$_{\rm LSR}$} & \colhead{$\Delta$RA} &
\colhead{$\Delta$Dec} & \colhead{$\mu_{\rm RA}$} &
\colhead{$ \mu_{\rm Dec}$}  \nl
\colhead{}  & \colhead{(km\,s$^{-1}$)} & \colhead{(mas)} & \colhead{(mas)} &
\colhead{(mas\,year$^{-1}$)} & \colhead{(mas\,year$^{-1}$)}}
\startdata
 1 & $-$70.6 & $-$40.654  & $-$61.958  & $-$0.19$\pm$0.01 & $-$0.20$\pm$0.01\\
 2 & $-$73.2 &     0.0    &     0.0    &    0.0           &    0.0          \\
 3 & $-$75.3 &     3.911  &  $-$0.152  & $-$0.08$\pm$0.03 &    0.20$\pm$0.01\\
 4 & $-$62.2 &    26.946  &   225.653  & $-$0.98$\pm$0.03 & $-$0.29$\pm$0.04\\
 5 & $-$66.0 &    26.794  &   226.021  & $-$0.90$\pm$0.05 & $-$0.16$\pm$0.11\\
 6 & $-$75.1 &    71.988  &   121.071  &    0.43$\pm$0.04 & $-$1.42$\pm$0.05\\
 7 & $-$74.6 &    72.074  &   121.087  &    0.04$\pm$0.07 & $-$1.24$\pm$0.07\\
 8 & $-$74.8 &    72.715  &   120.628  & $-$0.02$\pm$0.03 & $-$1.19$\pm$0.08\\
 9 & $-$71.3 & $-$41.526  & $-$63.426  & $-$0.05$\pm$0.09 &    0.04$\pm$0.09\\
10 & $-$73.6 &     7.187  &  $-$2.340  &    0.66$\pm$0.01 &    0.01$\pm$0.03\\
11 & $-$73.8 &     7.062  &  $-$2.330  &    0.66$\pm$0.01 & $-$0.01$\pm$0.03\\
12 & $-$75.5 &     5.558  &  $-$0.486  &    0.17$\pm$0.01 &    0.43$\pm$0.05\\
\hline
\multicolumn{2}{l}{Average} & &   & $-$0.02$\pm$0.15 (0.5) & $-$0.35$\pm$0.18 (0.6)  \nl
\hline
\enddata
\end{deluxetable}


\begin{thebibliography}{}
\bibitem[Benjamin et al. (2005)]{benja05}
{Benjamin, R. A., Churchwell, E., Babler, B. L. et al.}, 2005, \apj, 630, 149
\bibitem[Benjamin (2008)]{benja08}
{Banjamin, R. A.}, 2008, ASP Conference Series, Vol. 387, p. 375
\bibitem[Brand \& Blitz(1993)]{bb93}
{Brand, J. and  Blitz, L.}, 1993, \aap, 275, 67
\bibitem[Brand et al. (2001)]{brand01}
{Brand, J., Wouterloot, J. G. A., Rudolph, A. L. and de Geus, E. J.}, 2001,
\aap, 377, 644
\bibitem[Brunthaler et al. 2005]{andreas05}
{Brunthaler, A., Reid, M. J. and  Falcke, H.}, 2005,
Future Directions in High Resolution Astronomy: The 10th Anniversary of the
VLBA, ASP Conference Proceedings, Vol. 340. Edited by J. Romney and M. Reid.,
455
\bibitem[Dehnen \& Binney 1998]{db98}
{Dehnen, W. and Binney, J. J.}, 1998, \mnras, 298, 387
\bibitem[Digel et al. 1996]{digel96}
{Digel, S. W., Lyder, D. A., Philbrick, A. J. et al.}, 1996,
\apj, 458, 561
\bibitem[Fey et al. (2004)]{fey04}
{Fey, A. L., Ma, C., Arias, E. F. et al.}, 2004, \aj, 127, 3587
\bibitem[Georgelin \& Georgelin (1976)]{gg76}
{Georgelin, Y. M.\& Georgelin, Y. P.}, 1976, \aap, 49, 57
\bibitem[G\'omez (2006)]{gomez06}{G\'omez, G. C.}, 2006, \aj, 132, 2376
\bibitem[Hachisuka et al.(2006)]{hachi06}
{Hachisuka, K., Brunthaler, A., Menten, K. M., Reid, M. J. et al.}, 2006,
\apj, 645, 337
\bibitem[Honma et al. (2007)]{honma07}{Honma, M., Bushimata, T., Choi, Y. K.
et al.}, 2007, \pasj, 59, 889
\bibitem[Humphreys (1978)]{humphreys78}{Humphreys, R. M.}, 1978, \apjs, 38, 309
\bibitem[Levine et al. 2006]{levine06}
{Levine, E. S., Blitz, L. and Heiles, C.}, 2006, Science, 312, 5781
\bibitem[Nakanishi \& Sofue(2003)]{nakanishi03}
{Nakanishi, H. and Sofue, Y.}, 2003, \pasj, 55, 191
\bibitem[Nakanishi \& Sofue(2006)]{nakanishi06}
{Nakanishi, H. and Sofue, Y.}, 2006, \pasj, 58, 847
\bibitem[Pandey et al. (2006)]{pandey06}{Pandey, A. K., Sharma, S. and
Ogura, K.}, 2006, \mnras, 373, 255
\bibitem[Pradel et al. 2006]{pradel06}{Pradel, N., Charlot, P. and Lestrade,
J.-F.}, 2006, \aap, 452, 1099
\bibitem[Reid(1993)]{reid93}{Reid, M. J.}, 1993, \araa, 31, 345
\bibitem[Reid \& Brunthaler 2004]{rb04}
{Reid, M. J. and  Brunthaler, A.}, 2004, \apj, 616, 872
\bibitem[Rudolph et al. (1996)]{rudolph96}
{Rudolph, Alexander L., Brand, J., de Geus, E. J. and  Wouterloot, J. G. A.},
1996, \apj, 458, 653
\bibitem[Russeil (2003)]{russeil03}{Russeil, D.}, 2003, \aap, 397, 133
\bibitem[Russeil et al. (2007)]{russeil07}{Russeil, D., Adami, C. and
Georgelin, Y. M.}, 2007, \aap, 470, 161
\bibitem[Wouterloot et al (1993)]{wb89}
{Wouterloot, J. G. A., Brand, J. and Fiegle, K.}, 1993, A\&AS, 98, 589
\bibitem[Wouterloot et al. (1995)]{wouterloot95}
{Wouterloot, J. G. A., Fiegle, K., Brand, J. and  Winnewisser, G.}, 1995,
\aap, 301, 236
\bibitem[Xu et al (2006)]{xu06}
{Xu, Y., Reid, M. J., Zheng, X. W. and  Menten, K. M.}, 2006,
\textit{Science}, 311, 54

\end{thebibliography}
\end{document}